\documentclass[11pt,a4paper]{article}

\usepackage{amsmath,amsfonts,graphicx,natbib,psfrag,color}
\usepackage[textsize=small]{todonotes}
\usepackage{algorithm}
\usepackage{booktabs}

\title{Efficient SMC$^2$ schemes for stochastic kinetic models}
\author{Andrew Golightly\thanks{email: \texttt{andrew.golightly@ncl.ac.uk}} \and\ Theodore Kypraios\thanks{email: \texttt{theodore.kypraios@nottingham.ac.uk}}}
\date{$^*$School of Mathematics \& Statistics, Newcastle University,\\
  Newcastle upon Tyne, NE1 7RU, UK\\ 
$^\dagger$School of Mathematical Sciences, University of Nottingham,\\
Nottingham, NG27 2RD, UK}

\begin{document}
\maketitle
\begin{abstract}
Fitting stochastic kinetic models represented by Markov jump processes within 
the Bayesian paradigm is complicated by the intractability of the observed 
data likelihood. There has therefore been considerable attention given to the 
design of pseudo-marginal Markov chain Monte Carlo algorithms for such models. However, these 
methods are typically computationally intensive, often require careful tuning 
and must be restarted from scratch upon receipt of new observations. Sequential 
Monte Carlo (SMC) methods on the other hand aim to efficiently reuse posterior samples 
at each time point. Despite their appeal, applying SMC schemes in scenarios with both 
dynamic states and static parameters is made difficult by the problem of 
particle degeneracy. A principled approach 
for overcoming this problem is to move each parameter particle through a Metropolis-Hastings 
kernel that leaves the target invariant. This rejuvenation step is key to a recently 
proposed SMC$^2$ algorithm, which can be seen as the pseudo-marginal analogue of 
an idealised scheme known as iterated batch importance sampling. Computing the parameter weights in 
SMC$^2$ requires running a particle filter over dynamic states to unbiasedly estimate the intractable 
observed-data likelihood up to the current time point. In this paper, we propose to use an 
auxiliary particle filter inside the SMC$^2$ scheme. Our method uses two recently 
proposed constructs for sampling conditioned jump processes and we find that the resulting inference 
schemes typically require fewer state particles than when using a simple bootstrap filter. 
Using two applications, we compare the performance of the proposed approach with 
various competing methods, including two global MCMC schemes.
\end{abstract}


\section{Introduction}
\label{sec:intro}

Markov jump processes (MJPs) are routinely used to describe the dynamics of discrete-valued 
processes evolving continuously in time. Application areas include (but are not limited to) 
systems biology \citep{GoliWilk05,Wilkinson06}, predator-prey interaction \citep{ferm2008,BWK08} 
and epidemiology \citep{lin2013b,mckinley2014}. Here, we focus on the MJP 
representation of a stochastic kinetic model (SKM), whereby transitions of species in a reaction 
network are described probabilistically via an instantaneous reaction rate or hazard, which depends 
on the current system state and a set of rate constants, with the latter typically the object of inference. 

Owing to the intractability of the observed-data likelihood, Bayesian inference for SKMs is typically 
performed via Markov chain Monte Carlo (MCMC). Early attempts based on data augmentation were used 
by \cite{gibson1998} (see also \cite{oneill1999}) in the context of epidemiology, and by 
\cite{BWK08} for more general reaction networks. Unfortunately, such methods 
can suffer from poor mixing due to dependence between the parameters 
and latent states to be imputed. Recently proposed pseudo-marginal MCMC schemes e.g. particle MCMC 
(pMCMC) \citep{andrieu10}, offer a promising alternative and have been successfully applied in both the 
epidemiology \citep{mckinley2014} and systems biology \citep{GoliWilk15} literature. However, 
these `global' inference schemes require careful selection and tuning of proposal mechanisms and 
must be restarted from scratch upon receipt of new observations or when assimilating information 
from multiple data sets. Moreover, the efficiency of such schemes depends heavily on the mechanism used to update the 
latent jump process. 

We therefore consider sequential Monte Carlo (SMC) schemes which recycle posterior samples 
from one time-point to the next through simple reweighting and resampling steps (see e.g. \cite{doucet01} 
for an introduction and \cite{jacob2015} for a recent review). The main drawback 
of SMC in scenarios with both dynamic states and static parameters is particle degeneracy: that is, 
when the number of distinct particles decreases over time. \emph{Ad-hoc} approaches for 
overcoming this problem include the use of jittering each static parameter particle 
before propagation to the next time-point \citep{gordon93,liu2001}. In special cases when 
the distribution of parameters given all latent states is tractable, this structure can be 
exploited to give a particle filter that uses conditional sufficient 
statistics to rejuvenate parameter samples \citep{storvik2002,fearnhead2002}. A related approach 
is the particle learning (PL) method of \citep{carvalho2010} which combines the use of 
conditional sufficient statistics with an auxiliary particle filter \citep{pitt1999}. 
As discussed in \cite{chopin2010} however, PL does not completely overcome the degeneracy issue. 
\cite{chopin2002} proposed a particle filter for static models (the so called iterated batch 
importance sampling (IBIS) algorithm) that weights parameter particles by the observed-data 
likelihood contributions at each time point. Particle degeneracy is mitigated via a 
resample-move step \citep{gilks2001}, which `moves' each parameter particle through a Metropolis-Hastings 
kernel that leaves the target invariant. This step can be executed subject to the fulfilment of some 
degeneracy criterion e.g. small effective sample size. Unfortunately, intractability of the observed 
data likelihood precludes the use of IBIS for the class of models considered here. 

The focus of this paper, therefore, is on the pseudo-marginal analogue of IBIS, which replaces 
the idealised particle weights with estimates obtained by running an SMC scheme 
over dynamic states for each parameter particle. The nested use of particle filters in this way 
results in an algorithm known as SMC$^2$ \citep{chopin2013}. The resample-move step is accomplished by moving each 
parameter particle through a pMCMC kernel. The algorithm allows for choosing the number 
of state particles dynamically, by monitoring the acceptance rate of the resample-move step. 
Furthermore, the output of the algorithm can be used to estimate the model evidence at virtually 
no additional computational cost. This feature is particularly useful in the context of model 
selection, for example, when choosing between competing reaction networks based on a given 
data set.
 
The simplest implementation of SMC$^2$ uses a bootstrap filter over dynamic states in both the 
reweighting and move steps. However, this is likely to be particularly inefficient 
unless the noise in the measurement error process dominates the intrinsic stochasticity in the MJP. 
In this case, highly variable estimates of the observed-data likelihood will lead to small 
effective sample sizes, increasing the rate at which the resample-move step is triggered. Moreover, 
use of a bootstrap filter driven pMCMC kernel is also likely to be highly inefficient, requiring 
many state particles to maintain a reasonable acceptance rate. In the special case of no measurement 
error, \cite{drovandi2016} use the alive particle filter of \cite{delmoral2015} to drive an SMC$^2$ scheme.

Our contribution is the development of an auxiliary particle filter for use 
inside the SMC$^2$ scheme. Our method uses two recently 
proposed constructs for sampling conditioned jump processes, and can be applied in scenarios 
when only observations on a subset of system components are available. Moreover, observations 
may be subject to additive Gaussian error. We find that the proposed approach typically requires 
fewer state particles than when using a simple bootstrap filter. 
Using two applications and both real and synthetic data, we compare the performance of the proposed approach with 
various competing methods, including alive SMC$^2$.

The remainder of this paper is organised as follows. In Section~2, a brief review of 
the Markov process representation of a reaction network is presented. Section~3 outlines 
the structure of the problem before presenting details of the auxiliary particle filter 
and its use inside SMC$^2$. The methodology is used in a 
number of applications in Section~5 before conclusions are drawn 
in Section~6.

\section{Stochastic kinetic models}
\label{sec:stochkin}
We give here a brief introduction to stochastic kinetic models and refer the reader to \cite{Wilkinson06} for 
an in-depth treatment.

Consider a reaction network involving $u$ species $\mathcal{X}_1,\linebreak[1]
\mathcal{X}_2,\ldots,\mathcal{X}_u$ and $v$ reactions $\mathcal{R}_1,\mathcal{R}_2,\ldots,\mathcal{R}_v$, 
with each reaction denoted by $\mathcal{R}_i$ and written as
\begin{align*}
 \mathcal{R}_i:\quad p_{i1}\mathcal{X}_1+p_{i2}\mathcal{X}_2+\cdots+p_{iu}\mathcal{X}_u &\longrightarrow q_{i1}\mathcal{X}_1+q_{i2}\mathcal{X}_2+\cdots+q_{iu}\mathcal{X}_u 
\end{align*}
where stoichiometric coefficients $p_{ij}$ and $q_{ij}$ are non-negative integers. 
When a type $i$ reaction does occur, the system state changes discretely, 
via the $i$th row of the so called net effect matrix $A$, a $v\times u$ matrix with $(i,j)$th element given by 
$q_{ij}-p_{ij}$. In what follows, for notational convenience, we work with the 
stoichiometry matrix defined as $S=A'$. Let $X_{j,t}$ denote the (discrete) number of species $\mathcal{X}_j$ at time
$t$, and let $X_t$ be the $u$-vector $X_t = (X_{1,t},X_{2,t},\linebreak[1] \ldots,\linebreak[0] X_{u,t})'$. 
The time evolution of $X_t$ can be described by a vector of rates (or hazards) of the reactions together 
with the stoichiometry matrix which describes the effect of each reaction on the state. We therefore 
define a rate function $h_i(X_t,c_i)$, giving the overall hazard of a type $i$ reaction
occurring, and we let this depend explicitly on the reaction rate constant $c_i$, as well 
as the state of the system at time $t$. We model the system with a Markov jump process (MJP), 
so that for an infinitesimal time increment $dt$, the probability of a type $i$ reaction 
occurring in the time interval $(t,t+dt]$ is $h_i(X_t,c_i)dt$.  Under the standard assumption of 
mass action kinetics, the hazard function for a particular 
reaction of type $i$ takes the form of the rate constant multiplied 
by a product of binomial coefficients expressing the number of ways in which 
the reaction can occur, that is
\[
h_i(X_t,c_i) = c_i\prod_{j=1}^u \binom{X_{j,t}}{p_{ij}}.
\] 
Values for $c=(c_1,c_2,\ldots,c_v)'$ and the initial system state $X_0=x_0$
complete specification of the Markov process. Although the transition probability 
associated with this process is rarely analytically tractable (except in some simple 
cases) generating exact realisations of the MJP is straightforward. This is due to the
fact that if the current time and state of the system are $t$ and
$X_t$ respectively, then the time to the next event will be
exponential with rate parameter
\[
h_0(X_t,c)=\sum_{i=1}^v h_i(X_t,c_i),
\]
and the event will be a reaction of type $\mathcal{R}_i$ with probability
$h_i(X_t,c_i)/h_0(X_t,c)$ independently of the inter-event time. This simulation 
method is typically referred to as \emph{Gillespie's direct method} in the stochastic kinetics
literature, after \cite{Gillespie77}.

\subsection{Example 1: a stochastic epidemic model}\label{ex1}
The Susceptible--Infected--Removed (SIR) epidemic model \citep[see e.g.][]{AnBr00} describes the evolution of two species (susceptibles $\mathcal{X}_{1}$ and infectives $\mathcal{X}_{2}$) via two reaction channels which correspond to an infection of a susceptible individual and a removal of an infective individual:
\begin{align*}
\mathcal{R}_1:\quad \mathcal{X}_{1}+\mathcal{X}_{2} &\longrightarrow  2\mathcal{X}_{2}\\
\mathcal{R}_2:\quad \mathcal{X}_{2} &\longrightarrow \emptyset.
\end{align*}
The stoichiometry matrix is given by
\[
S = \left(\begin{array}{rr} 
-1 & 0\\
 1 & -1
\end{array}\right)
\]
and the associated hazard function is 
\[
h(X_t,c) = (c_{1} X_{1,t}X_{2,t}, c_{2} X_{2,t})'.
\]
\subsection{Example 2: prokaryotic autoregulation}\label{ex2}
A commonly used mechanism for
auto-regulation in prokaryotes which has been well-studied and
modelled is a negative feedback mechanism whereby dimers of a protein
repress its own transcription \citep[e.g.][]{ARM98}. A 
simplified model for such a prokaryotic auto-regulation, 
based on this mechanism of dimers of a protein coded 
for by a gene repressing its own transcription, can be 
found in \cite{GoliWilk05} \cite[see also][]{GoliWilk11}. 
The full set of reactions in this simplified model are
\begin{align*}
\mathcal{R}_1:\quad \textsf{DNA}+\textsf{P}_2 &\longrightarrow
\textsf{DNA}\cdot\textsf{P}_2 \\
\mathcal{R}_2:\quad \textsf{DNA}\cdot\textsf{P}_2 &\longrightarrow
\textsf{DNA}+\textsf{P}_2 \\
\mathcal{R}_3:\quad \textsf{DNA} &\longrightarrow \textsf{DNA} + \textsf{RNA} \\
\mathcal{R}_4:\quad \textsf{RNA} &\longrightarrow \textsf{RNA} + \textsf{P} \\
\mathcal{R}_5:\quad 2\textsf{P} &\longrightarrow \textsf{P}_2 \\
\mathcal{R}_6:\quad \textsf{P}_2 &\longrightarrow 2\textsf{P} \\
\mathcal{R}_7:\quad \textsf{RNA} &\longrightarrow \emptyset \\
\mathcal{R}_8:\quad \textsf{P} &\longrightarrow \emptyset.
\end{align*}
Note that this model contains a conservation law, so that the total 
number $k$ of $\textsf{DNA}\cdot\textsf{P}_2$ and $\textsf{DNA}$ 
is fixed for all time. Denoting the number of molecules of $\textsf{RNA}$, 
$\textsf{P}$, $\textsf{P}_2$ and $\textsf{DNA}$ as $X_1$, $X_2$, $X_3$ and 
$X_4$ respectively, gives the stoichiometry matrix 
\[
S = \left(\begin{array}{rrrrrrrr}
0&\phantom{-}0&\phantom{-}1&\phantom{-}0&0&0&-1&0\\
0&0&0&1&-2&2&0&-1\\
-1&1&0&0&1&-1&0&0\\
-1&1&0&0&0&0&0&0\\
\end{array}\right),
\]
and associated hazard function
\begin{align*}
h(X,c) &= (c_1 X_{4}X_{3}, c_2(k-X_{4}),c_3 X_{4}, c_4 X_{1},c_5 X_{2}(X_{2}-1)/2,c_6 X_{3}, c_7 X_{1}, c_8 X_{2})'.
\end{align*}
where we have dropped $t$ to ease the notation.

\section{Sequential Bayesian inference}
\label{sec:bayes}

\subsection{Setup}

Suppose that the Markov jump process is not observed directly, 
but observations (on a regular grid) $y_{t},t=1,2,\ldots$ are available 
and assumed conditionally independent (given the latent jump process) with conditional probability 
distribution obtained via the observation equation
\begin{equation}\label{obs_eq}
Y_{t}=P'X_{t}+\varepsilon_{t},\qquad \varepsilon_{t}\sim \textrm{N}\left(0,\Sigma\right),\qquad t=1,2,\ldots
\end{equation} 
Here, $Y_{t}$ is taken to be a length-$p$ vector, $P$ is a constant matrix of dimension 
$u\times p$ and $\varepsilon_{t}$ is a length-$p$ 
Gaussian random vector. The density $p(y_{t}|x_{t})$ linking the observed and latent 
processes satisfies
\[
p(y_{t}|y_{1:t-1},x_{[1,t]},c)=p(y_{t}|x_{t},c)
\] 
where $x_{[1,t-1]}$ denotes the MJP over an interval $[1,t-1]$. 

We assume that 
primary interest lies in the recursive exploration of the marginal posteriors 
$p(c|y_{1:t})$, $t=1,\ldots,T$. Upon ascribing a prior density $p(c)$ to the 
parameters, Bayes theorem gives 
\begin{align}
p(c|y_{1:t})&\propto p(c)p(y_{1:t}|c)  \\
&\propto p(c|y_{1:t-1})p(y_{t}|y_{1:t-1},c) \label{post}
\end{align}
which immediately suggests a sequential importance sampling scheme that repeatedly 
reweights a set of $N_c$ parameter samples (known as `particles' in this context) by the observed-data (or `marginal') 
likelihood contributions $p(y_{t}|y_{1:t-1},c)$. This approach is used in the iterated batch importance 
sampling (IBIS) algorithm of \cite{chopin2002}, together with MCMC steps for 
rejuvenating parameter samples in order to circumvent particle degeneracy. Although 
each observed-data likelihood contribution is typically intractable, progress can be made by substituting 
a non-negative estimate of $p(y_{t}|y_{1:t-1},c)$. In order for the resulting algorithm 
to target the correct posterior, these estimates should be constructed 
so that the observed-data likelihood up to the current time point, $p(y_{1:t}|c)$, 
can be unbiasedly estimated. This task can be achieved 
by running a particle filter with $N_{x}$ particles targeting $p(x_{t}|y_{1:t},c)$ for each 
$c$-particle. Particle MCMC steps are then occasionally used to rejuvenate the sample. This 
approach was proposed and theoretically justified by \cite{chopin2013} 
who term the resulting algorithm \emph{SMC$^2$} due to the use of nested filters. The simplest 
implementation of the algorithm runs a bootstrap particle filter \citep[e.g.][]{gordon93} 
for each $c$-particle, which only requires the ability to forward-simulate the MJP and evaluate 
$p(y_{t}|x_{t},c)$. Despite the appeal of this simple approach, the resulting estimates of 
the observed-data likelihood contributions can have high variance, unless the observations are 
not particularly informative, limiting the efficiency of the SMC$^2$ scheme. This is due to 
the collapse of the bootstrap particle filter, which results from very few state trajectories 
having reasonable weight. The problem is exacerbated in the case of no measurement error, where 
only state trajectories that `hit' observations are assigned a non-zero weight. \cite{drovandi2016} use the alive 
particle filter of \cite{delmoral2015} (see also Appendix~\ref{alive}) to avoid this problem. Unfortunately, this approach can be 
extremely computationally expensive, since it repeatedly generates simulations of the jump process 
until a predetermined number of hits are obtained. In what follows, therefore, we use an auxiliary particle filter (for which the 
bootstrap filter can viewed as a special case) to efficiently estimate each $p(y_{t}|y_{1:t-1},c)$. 
We describe the auxiliary particle filter in the next section before describing its use inside 
an SMC$^2$ scheme.

\subsection{Auxiliary particle filter}\label{pf}

The aim of the particle filter is to recursively approximate the sequence of filtering 
densities $p(x_{t}|y_{1:t},c)$. To this end, suppose that at time $t-1$, a weighted 
sample $\{x_{t-1}^i,w_{t-1,c}^i\}_{i=1}^{N_x}$ is available, and is approximately 
distributed according to $p(x_{t-1}|y_{1:t-1},c)$. Note that although the predictive 
$p(x_{(t-1,t]}|y_{1:t-1},c)$ is typically intractable, the weighted sample from the 
previous time point can be used to give the approximation 
$\hat{p}(x_{(t-1,t]}|y_{1:t-1},c)\propto \sum_{i=1}^{N_x}p(x_{(t-1,t]}|x_{t-1}^{i},c)w_{t-1,c}^i$. 
Hence, upon receipt of a new datum $y_t$, the particle filter constructs the approximate 
posterior
\begin{equation}\label{pftarget}
\hat{p}(x_{(t-1,t]}|y_{1:t},c)\propto p(y_{t}|x_{t},c)\sum_{i=1}^{N_x}p(x_{(t-1,t]}|x_{t-1}^{i},c)w_{t-1,c}^i
\end{equation}
from which draws can be generated using (for example) importance resampling. A simple strategy 
is to use $\hat{p}(x_{(t-1,t]}|y_{1:t-1},c)$ as a proposal mechanism, which is straightforward to sample from 
by picking a particle $x_{t-1}^i$ with probability $w_{t-1,c}^i$ and simulating according to $p(x_{(t-1,t]}|x_{t-1}^{i},c)$ 
using Gillespie's direct method (see Section~\ref{sec:stochkin}). The state $x_{t}^i$ can be stored along with the new (unnormalized) weight 
$\tilde{w}_{t,c}^i = p(y_{t}|x_{t}^i,c)$. Resampling (with replacement) amongst the particles using the weights as probabilities 
gives a sample approximately distributed according to (\ref{pftarget}). Repeating this procedure for each time point gives the bootstrap particle filter 
of \cite{gordon93}.    

The auxiliary particle filter (APF) of \cite{pitt1999} \citep[see also][]{pitt12} can be seen as a generalization of the 
bootstrap filter. The APF is constructed by noting that
\begin{align*}
p(y_{t}|x_{t},c)p(x_{(t-1,t]}|x_{t-1},c)&=p(y_{t}|x_{t-1},c)p(x_{(t-1,t]}|x_{t-1},y_{t},c)
\end{align*}
which immediately suggests an importance resampling strategy that initially preweights each $x_{t-1}^i$ 
particle by $\tilde{w}_{t-1|t,c}=p(y_{t}|x_{t-1}^i,c)w_{t-1,c}^i$ and propagates according to $p(x_{(t-1,t]}|x_{t-1}^i,y_{t},c)$. 
The new (unnormalized) weight is $\tilde{w}_{t,c}^i = 1$, giving the fully adapted form of the APF \citep{pitt2001}. In practice, 
$p(y_{t}|x_{t-1},c)$ and $p(x_{(t-1,t]}|x_{t-1},y_{t},c)$ are intractable and approximations 
$g(y_{t}|x_{t-1},c)$ and $g(x_{(t-1,t]}|x_{t-1},y_{t},c)$ must be sought, giving the APF described in 
Algorithm~\ref{auxPF}. Note that taking $g(y_{t}|x_{t-1},c)=1$ and $g(x_{(t-1,t]}|x_{t-1},y_{t},c)=p(x_{(t-1,t]}|x_{t-1},c)$ 
admits the bootstrap particle filter as a special case.

Following \cite{pitt12}, we use the output of the APF to estimate $p(y_{t}|y_{1:t-1},c)$ with the quantity
\[
\hat{p}(y_{t}|y_{1:t-1},c)=\left(\sum_{i=1}^{N_{x}}\frac{\tilde{w}_{t,c}^{i}}{N_{x}}\right)\left(\sum_{i=1}^{N_{x}}\tilde{w}_{t-1|t,c}\right).
\]
Crucially, \cite{pitt12} \citep[see also][]{delmoral04} show that 
\[
\hat{p}(y_{1:T}|c)=\hat{p}(y_{1}|c)\prod_{t=2}^{T}\hat{p}(y_{t}|y_{1:t-1},c)
\] 
is an unbiased estimator of $p(y_{1:T}|c)$. Justification of the use of $\hat{p}(y_{t}|y_{1:t-1},c)$, as given above, 
in an SMC$^2$ scheme then follows directly from \cite{chopin2013}.

\begin{algorithm}[t]
\caption{Auxiliary particle filter}\label{auxPF}
\begin{enumerate}
\item Initialisation ($t=1$). For $i=1,2,\ldots ,N_x$:
\begin{itemize}
\item[(a)] Sample $x_{1}^{i}\sim p(\cdot)$.
\item[(b)] Compute the weights $\tilde{w}_{1,c}^{i}=p(y_{1}|x_{1}^{i},c)$, \ $w_{1,c}^{i}=\frac{\tilde{w}_{1,c}^{i}}{\sum_{j=1}^{N_{x}}\tilde{w}_{1,c}^{j}}$.
\item[(c)] Compute the current estimate of observed-data likelihood $\hat{p}(y_{1}|c)=\sum_{i=1}^{N_x}\tilde{w}_{1,c}^{i}/N_x$.
\end{itemize}
\item For times $t=2,3,\ldots ,T$ and $i=1,2,\ldots ,N_x$:
\begin{itemize}
\item[(a)] Compute the preweights $\tilde{w}_{t-1|t,c}^{i}=g(y_{t}|x_{t-1}^{i},c)w_{t-1,c}^{i}$, \ 
$w_{t-1|t,c}^{i}=\frac{\tilde{w}_{t-1|t,c}^{i}}{\sum_{i=1}^{N_{x}}\tilde{w}_{t-1|t,c}^{i}}$.
\item[(b)] Sample the index $a_{t-1}^{i}\sim \mathcal{M}\big(w_{t-1|t,c}^{1:N_{x}}\big)$ of the ancestor of particle $i$, where $\mathcal{M}(p^{1:n})$ denotes the 
multinomial distribution that assigns a probability $p^i$ to outcome $i\in \{1,\ldots,n\}$.
\item[(c)] Propagate. Sample $x_{(t-1,t]}^{i}\sim g\big(\cdot|x_{t-1}^{a_{t-1}^{i}},y_{t},c\big)$.
\item[(d)] Compute the weights
\[
\tilde{w}_{t,c}^{i}=\frac{p(y_{t}|x_{t}^{i},c)p\big(x_{(t-1,t]}^{i}|x_{t-1}^{a_{t-1}^{i}},c\big)}
{g\big(y_{t}|x_{t-1}^{a_{t-1}^{i}},c\big)g\big(x_{(t-1,t]}^{i}|x_{t-1}^{a_{t-1}^{i}},y_{t},c\big)}, 
\]
\[
w_{t,c}^{i}=\frac{\tilde{w}_{t,c}^{i}}{\sum_{j=1}^{N_{x}}\tilde{w}_{t,c}^{j}}
\]
\item[(e)] Compute the current estimate of observed-data likelihood $\hat{p}(y_{1:t}|c)=\hat{p}(y_{1:t-1}|c)\hat{p}(y_{t}|y_{1:t-1},c)$ where
\[
\hat{p}(y_{t}|y_{1:t-1},c)=\left(\sum_{i=1}^{N_{x}}\frac{\tilde{w}_{t,c}^{i}}{N_{x}}\right)\left(\sum_{i=1}^{N_{x}}\tilde{w}_{t-1|t,c}\right).
\]
\end{itemize}
\end{enumerate}
\end{algorithm}

\subsubsection{Propagation - method 1}
\label{sec:cond}

It remains that we can find suitable densities $g(y_{t}|x_{t-1},c)$ and 
$g(x_{(t-1,t]}|x_{t-1},y_{t},c)$. Focusing first on the latter, we use an 
approximation to the conditioned jump process proposed by \cite{GoliWilk15}. 
The method works by approximating the expected number of reaction events over 
an interval of interest conditional on the next observation. The resulting conditioned 
hazard is used in place of the unconditioned hazard in Gillespie's direct method. 

Consider an interval $[t-1,t]$ and suppose that we have simulated as far as time 
$s\in[t-1,t]$. Let $\Delta R_{s}$ denote the number of 
reaction events over the time $t-s=\Delta s$. We approximate $\Delta R_{s}$ 
by assuming a constant reaction hazard over $\Delta s$. A Gaussian 
approximation to the corresponding Poisson distribution then gives
\[
\Delta R_{s}\sim \textrm{N}\left(h(x_s,c)\Delta s\,,\,H(x_s,c)\Delta s\right)
\]
where $H(x_s,c)=\textrm{diag}\{h(x_s,c)\}$. Under the Gaussian observation 
regime (\ref{obs_eq}) we have that
\begin{align}\label{preweight}
Y_{t}|X_{s}=x_{s} &\sim \textrm{N}\left(P'\left(x_{s}+S\,h(x_s,c)\Delta s \right)\,,\, P'S\,H(x_s,c)S'P\Delta s +\Sigma  \right).
\end{align}
Hence, the joint distribution of $\Delta R_{s}$ 
and $Y_{t}$ (conditional on $x_s$) can then be obtained approximately as
\begin{align*}
 \begin{pmatrix} \Delta R_{s} \\ Y_{t} \end{pmatrix}
&\sim \textrm{N}\left\{\begin{pmatrix} h(x_s,c)\Delta s \\ P'\left(x_{s}+S\,h(x_s,c)\Delta s\right)\end{pmatrix}\,,\, 
\begin{pmatrix} H(x_s,c)\Delta s & H(x_s,c)S'P\Delta s\\
P'S\,H(x_s,c)\Delta s & P'S\,H(x_s,c)S'P\Delta s +\Sigma\end{pmatrix}\right\}.
\end{align*}
Taking the expectation of $\Delta R_{s}|Y_{t}=y_{t}$ 
and dividing the resulting expression by $\Delta s$ gives 
an approximate conditioned hazard as
\begin{align}
h^{*}(x_s,c|y_{t})&=h(x_s,c)\nonumber \\
&\quad+H(x_s,c)S'P\left(P'S\,H(x_s,c)S'P\Delta s +\Sigma\right)^{-1}\left(y_{t}-P'\left[x_{s}+S\,h(x_s,c)\Delta s\right]\right). \label{haz}
\end{align}
Although the conditioned hazard in (\ref{haz}) depends on the current time $s$ 
in a nonlinear way, a simple implementation ignores 
this time dependence, giving exponential waiting times between reaction events. 
Hence, the construct can be used to generate realisations from an approximation to 
the true (but intractable) conditioned jump process by applying Gillespie's direct method 
with $h(x_s,c)$ replaced by $h^{*}(x_s,c|y_{t})$. 

To calculate the weights used in step 2(d) of Algorithm~\ref{auxPF}, we note 
that $p(x_{(t-1,t]}|x_{t-1},c)$ can be 
written explicitly by considering the generation of all reaction times and types over 
$(t-1,t]$. To this end, we let $r_{j}$ denote the number of reaction events of type $\mathcal{R}_{j}$, 
$j=1,\ldots,v$, and define $n_{r}=\sum_{j=1}^{v}r_{j}$ as the total number of reaction events 
over the interval, which is obtained deterministically from the trajectory $x_{(t-1,t]}$. 
Reaction times (assumed to be in increasing order) and types are denoted by 
$(\tau_{i},\nu_{i})$, $i=1,\ldots ,n_{r}$, $\nu_{i}\in \{1,\ldots ,v\}$ and we take $\tau_{0}=t-1$ and 
$\tau_{n_{r}+1}=t$. The so-called complete-data likelihood \citep{Wilkinson06} over $(t-1,t]$ is then 
given by
\begin{align*}
p(x_{(t-1,t]}|x_{t-1},c) &=\left\{\prod_{i=1}^{n_{r}}h_{\nu_{i}}\left(x_{\tau_{i-1}},c_{\nu_{i}}\right)\right\}\exp\left\{-\sum_{i=1}^{n_{r}}h_{0}\left(x_{\tau_{i}},c\right)\left(\tau_{i+1}-\tau_{i}\right)\right\}
\end{align*} 
An expression for $g(x_{(t-1,t]}|x_{t-1},y_{t},c)$ is obtained similarly. Hence the 
weights we require take the form 
\begin{align}
\tilde{w}_{t,c}&=\frac{p(y_{t}|x_{t},c)p(x_{(t-1,t]}|x_{t-1},c)}{g(y_{t}|x_{t-1},c)g(x_{(t-1,t]}|x_{t-1},y_{t},c)}\nonumber \\
&=\frac{p(y_{t}|x_{t},c)}{g(y_{t}|x_{t-1},c)}\left\{\prod_{i=1}^{n_{r}}\frac{h_{\nu_{i}}\left(x_{\tau_{i-1}},c_{\nu_{i}}\right)}{h^{*}_{\nu_{i}}\left(x_{\tau_{i-1}},c_{\nu_{i}}|y_{t}\right)}\right\}\exp\left\{-\sum_{i=1}^{n_{r}}\left[h_{0}\left(x_{\tau_{i}},c\right)-h^{*}_{0}\left(x_{\tau_{i}},c|y_{t}\right)\right]\left[\tau_{i+1}-\tau_{i}\right]\right\}. \label{weight}
\end{align}

\subsubsection{Propagation - method 2}
\label{sec:cond2}

\cite{fearnhead2008} derives a conditioned hazard in the case of complete and noise-free observation of the 
MJP. Extending the method to the observation scenario given by (\ref{obs_eq}) is straightforward. 
Consider again an interval $[t-1,t]$ and suppose that we have simulated as far as time 
$s\in[t-1,t]$. For reaction $\mathcal{R}_i$ let $x'=x_{s}+S^{(i)}$, where $S^{(i)}$ denotes the $i$th column of the stoichiometry 
matrix so that $x'$ is the state of the MJP after a single occurrence of $\mathcal{R}_i$. The conditioned 
hazard of $\mathcal{R}_i$ satisfies
\begin{align}
h_{i}(x_s,c|y_t)&=\lim_{\Delta s\to 0}\frac{Pr(X_{s+\Delta s}=x'|X_{s}=x_{s},y_{t},c)}{\Delta s} \nonumber \\
&=h_{i}(x_s,c_i)\lim_{\Delta s\to 0}\frac{p(y_{t}|X_{s+\Delta s}=x',c)}{p(y_{t}|X_{s}=x_s,c)} \nonumber \\
&=h_{i}(x_s,c_i)\frac{p(y_{t}|X_{s}=x',c)}{p(y_{t}|X_{s}=x_s,c)}. \nonumber
\end{align}
Of course in practice, the transition density $p(y_{t}|x_{s},c)$ is intractable and we therefore use 
the approximation in (\ref{preweight}) to obtain an approximate conditioned hazard $h_{i}^{\dagger}(x_s,c|y_t)$ 
and combined hazard $h_{0}^{\dagger}(x_s,c|y_t)$. Note that to calculate this approximate 
conditioned hazard, the density associated with the approximation in (\ref{preweight}) must be calculated 
$v+1$ times (once using $x_s$ and for each $x'$ obtained after the $v$ possible transitions of the process). 
Although $h_{0}^{\dagger}(x_s,c|y_t)$ is time dependent, the simple simulation approach described in Section~\ref{sec:cond} 
that ignores this time dependence can be easily implemented. The form of the weight required in step 2(d) 
of Algorithm~\ref{auxPF} is given by equation~\ref{weight} with $h^*$ replaced by $h^\dagger$.

\subsubsection{Preweight}

Finally, note that the derivations of the 
conditioned hazards described above suggest a form for the preweight $g(y_{t}|x_{t-1},c)$. Using the 
approximation in (\ref{preweight}) with $s=t-1$ and assuming an inter-observation time 
of $\Delta$ gives
\begin{align}\label{preweight2}
g(y_{t}|x_{t-1},c)&=N\left(y_{t}; P'\left(x_{t-1}+S\,h(x_{t-1},c)\Delta \right)\,, \, P'S\,H(x_{t-1},c)S'P\Delta +\Sigma\right)
\end{align} 
where $N(\cdot;m,V)$ denotes the multivariate Gaussian density with mean vector $m$ and variance matrix $V$. 
In some scenarios, the density in (\ref{preweight2}) may have lighter tails 
than $p(y_{t}|x_{t-1},c)$. In this case, some particles that are consistent with the next observation are likely 
to be pruned out. Although the problem can be alleviated by raising the density in (\ref{preweight2}) to 
a power (say $1/\delta$ where $\delta>1$), this introduces an additional tuning parameter. We find that simply 
taking $g(y_{t}|x_{t-1},c)=1$ is computationally convenient and works well in practice.

\subsection{SMC$^2$ scheme}\label{smc2}

In this section, we provide a brief exposition of the SMC$^2$ scheme. The reader is referred to 
\cite{chopin2013} for further details including a formal justification 
(see also \cite{fulop2013} for a related algorithm and \cite{jacob2015} for 
a recent discussion). 

Recall the target posterior at time $t$, $p(c|y_{1:t})$ given by (\ref{post}). Suppose that a weighted sample 
$\{c^{k},\omega^{k}\}_{k=1}^{N_c}$ from $p(c|y_{1:t})$ is available. The SMC$^2$ algorithm reweights each $c$-particle 
according to a non-negative estimate of $p(y_{t}|y_{1:t-1},c^k)$, obtained from the output of a particle filter. 
We propose to use the auxiliary particle filter of Section~\ref{pf}. In order to use the APF in this way, we require 
storage of the state particles and associated weights at each time point $t$ and for each parameter particle $c^k$. 
We denote the APF output at iteration $t$ by $\{x_{t,c^{k}}^{1:N_x},w_{t,c^{k}}^{1:N_x}\}$. To circumvent particle 
degeneracy, the SMC$^2$ scheme uses a resample-move step \citep[see e.g.][]{gilks2001} that firstly resamples parameter 
particles (and the associated states, weights and observed-data likelihoods $p(y_{1:t}|c^{k})$) and then moves each 
parameter sample through a particle Metropolis-Hastings kernel which leaves the target posterior invariant \citep{andrieu10}. 
The resample-move step is only used if some degeneracy criterion is fulfilled. Typically, at each time $t$, the 
effective sample size (ESS) is computed as 
\[
\textrm{ESS}=1\big / \,\,{\sum_{k=1}^{N_c}(\omega^{k})^2}
\] 
and the resample-move step is triggered if $\textrm{ESS}<\gamma N_c$ for $\gamma \in(0,1)$ and a standard choice is 
$\gamma=0.5$. A key feature of the SMC$^2$ scheme is that the current set of $c$-particles can be used in the 
design of the proposal density $q(c^*|c)$. For the applications in Section~\ref{app}, we use an independent 
proposal so that $q(c^*|c)=q(c^*)$. As the rate constants must be strictly positive, we take
\[
q(c^*)=logN\left(c^*; \widehat{E}(\log(c)|y_{1:t}),\widehat{Var}(\log(c)|y_{1:t})\right)
\]
where $logN(\cdot;m,V)$ denotes the density associated with the exponential of a $N(m,V)$ random variable.

The SMC$^2$ scheme with fixed $N_x$ is given by Algorithm~\ref{smc2alg}. It remains that the number of state particles 
is suitably chosen. \cite{andrieu10} show that $N_x=O(t)$ to obtain a reasonable acceptance rate in the particle 
Metropolis-Hastings step. Therefore, \cite{chopin2013} suggest an automatic method that allows $N_x$ to increase over time. 
Essentially, the acceptance rate of the move step is monitored and if this rate falls below a given threshold, $N_x$ 
is increased (e.g. by multiplying by 2). Suppose that at time $t$ and for each $c^k$, we have 
$\{x_{t,c^{k}}^{1:N_x},w_{t,c^{k}}^{1:N_x}\}$ and observed-data likelihood $\hat{p}_{N_{x}}(y_{1:t}|c^k)$, where we have 
explicitly written the observed-data likelihood to depend on $N_x$. Let $\tilde{N}_x$ denote the updated number of state 
particles. A generalized importance sampling strategy is used to swap the $x$-particles, their associated weights and 
the estimates of observed-data likelihood with new values obtained by running the APF with $\tilde{N}_x$ state particles, for 
each $c^k$. \cite{chopin2013} show that the weights associated with each parameter particle $c^k$ should be 
multiplied by $\hat{p}_{N_{x}}(y_{1:t}|c^k)/ \hat{p}_{\tilde{N}_{x}}(y_{1:t}|c^k)$. Fortunately, the frequency at which 
the potentially expensive resample-move step is executed reduces over time and the computational cost of the algorithm 
is $O(N_c t^2)$ (rather than $O(N_c t^3)$ if the resample-move step was triggered at every time-point).

\begin{algorithm}[t]
\caption{SMC$^2$ scheme}\label{smc2alg}
\begin{enumerate}
\item Initialisation. For $k=1,\ldots,N_c$ sample $c^k\sim p(\cdot)$ and set $\tilde{\omega}^k=1$. For $t=1,\ldots ,T$:
\item Sequential importance sampling. For $k=1,\ldots,N_c$:
\begin{itemize}
\item[(a)] Perform iteration $t$ of the auxiliary particle filter to obtain $\{x_{t,c^{k}}^{1:N_x},w_{t,c^{k}}^{1:N_x}\}$ 
and $\hat{p}(y_{t}|y_{1:t-1},c^k)$. Note that $\hat{p}(y_1|c^k)=\hat{p}(y_{1}|y_{1:0},c^k)$. 
\item[(b)] Update and normalize the importance weights via
\[
\tilde{\omega}^k:=\tilde{\omega}^k\hat{p}(y_{t}|y_{1:t-1},c^k), \qquad \omega^{k}=\frac{\tilde{\omega}^{k}}{\sum_{j=1}^{N_{c}}\tilde{\omega}^{j}}
\]
\item[(c)] Update observed-data likelihood estimate via
\[
\hat{p}(y_{1:t}|c^k)=\hat{p}(y_{1:t-1}|c^k)\hat{p}(y_{t}|y_{1:t-1},c^k).
\]
\end{itemize}
\item If $\textrm{ESS}<\gamma N_c$ resample and move. For $k=1,\ldots,N_c$:
\begin{itemize}
\item[(a)] Sample indices $a_{k}\sim \mathcal{M}\big(\omega^{1:N_c}\big)$ 
and set $\{c^k,\tilde{\omega}^k\}:=\{c^{a_{k}},1\}$, $\{x_{t,c^k}^{1:N_x},w_{t,c^k}^{1:N_x}\}:=\{x_{t,c^{a_{k}}}^{1:N_x},w_{t,c^{a_{k}}}^{1:N_x}\}$ 
and $\hat{p}(y_{1:t}|c^{k}):=\hat{p}(y_{1:t}|c^{a_k})$.
\item[(b)] Propose $c^*\sim q(\cdot|c^k)$. Perform iterations $1,\ldots,t$ 
of the auxiliary particle filter to obtain $\hat{p}(y_{1:t}|c^{*})$. With probability
\[
\textrm{min}\left\{1, \frac{p(c^*)\hat{p}(y_{1:t}|c^{*})}{p(c^k)\hat{p}(y_{1:t}|c^{k})}\times \frac{q(c^k|c^*)}{q(c^*|c^k)}\right\}
\]
put $c^k:=c^*$, $\{x_{t,c^k}^{1:N_x},w_{t,c^k}^{1:N_x}\}:=\{x_{t,c^{*}}^{1:N_x},w_{t,c^{*}}^{1:N_x}\}$ and 
$\hat{p}(y_{1:t}|c^{k}):=\hat{p}(y_{1:t}|c^{*})$.
\end{itemize}
\end{enumerate}
\end{algorithm}

Finally, consider the evidence 
\[
p(y_{1:T})=\prod_{t=1}^{T}p(y_{t}|y_{1:t-1}), 
\]
where we adopt the convention that $p(y_1)=p(y_{1}|y_{1:0})$. It is straightforward to estimate $p(y_{1:T})$ using the output of the SMC$^2$ scheme, 
at virtually no additional computational cost. Each factor $L_t=p(y_{t}|y_{1:t-1})$ in the product above is estimated by
\begin{equation}\label{ev}
\hat{L}_t=\sum_{k=1}^{N_c}\omega^k \hat{p}(y_{t}|y_{1:t-1},c^k).
\end{equation}

\section{Applications}\label{app}

To illustrate the methodology described in the previous sections we consider two applications of increasing complexity. In Section~\ref{epi}, a Susceptible--Infected--Removed (SIR) epidemic model is fitted using real data; namely, the Abakaliki smallpox data set given in \cite{bailey1975}. We compare the performance of SMC$^2$ schemes  based on auxiliary, bootstrap and alive particle filters. Using synthetic data, we compare the best performing SMC$^2$ scheme with its particle MCMC counterpart and, additionally, a data augmentation scheme. In Section~\ref{prok}, we apply SMC$^2$ to infer the parameters governing a simple prokaryotic autoregulatory network using synthetic data. All algorithms are coded in C and were run on a desktop computer with an Intel Core i7-4770 processor and a 3.40GHz clock speed. The code is available at \emph{http://www.mas.ncl.ac.uk/$\sim$nag48/smc2.zip}.

\subsection{Abakaliki smallpox data}\label{epi}

We first consider the well-studied temporal data set obtained from a smallpox outbreak that took place in the small Nigerian village Abakaliki in 1967. \citet[p. 125]{bailey1975} provides a complete set of 29 inter-removal times, measured in days. Table \ref{tab:tabEpi} shows the data here as the days on which the removal of individuals actually took place, with the first day set to be time 0. The outbreak resulted in 32 cases, 30 out of which corresponded to individuals who were members of a religious organisation whose 120 members refused to be vaccinated. 

Numerous authors such as \cite{oneill1999}, \cite{fearnhead2004} and \cite{boys2007} among others have considered these data by focussing solely on the 30 cases among the population of 120, despite the fact that the original dataset (provided in a WHO report) contains far more information than the inter-removal times, such as the physical locations of the cases and the members of each household. A fully Bayesian analysis of this full dataset can be found in \cite{StocKypOn17} but here our purpose is to illustrate our methodology and therefore, we only consider the partial data set assuming that there have been 30 cases in a closed population of size 120. 

We assume an SIR model (see Section~\ref{ex1}) for the data with observations being 
equivalent to daily measurements of $X_{1}+X_{2}$ (as there is a fixed population size). In addition, 
and for simplicity, we assume that a single individual remained infective just after the first 
removal occurred. We analyse the data under the assumption of no measurement error, that is, $P'=(1,1)$ and 
$\Sigma=0$ in the observation equation (\ref{obs_eq}).

\begin{table*}[t]
\centering
\small
	\begin{tabular}{@{} lllllllllllllll@{}}

Day & 1 & 14 & 21 & 23 & 26 & 27 & 31 &36 &39 & 41 & 43 &48\\
No. of removals & 1 & \phantom{0}1 & \phantom{0}1 & \phantom{0}1 & \phantom{0}3  &\phantom{0}1 & \phantom{0}1 & \phantom{0}1 &\phantom{0}1 & \phantom{0}2 &\phantom{0}2 &\phantom{0}1\\ 
\midrule
Day & 51 & 52 & 56 & 57 & 58 & 59 & 61 &62 &67 & 72 & 77 & \\
No. of removals & \phantom{0}1 & \phantom{0}1 & \phantom{0}2 &\phantom{0}1 &\phantom{0}1 &\phantom{0}1 & \phantom{0}2 &\phantom{0}1 &\phantom{0}2 &\phantom{0}1 &\phantom{0}1 & \\ 

	\end{tabular}
	\caption{Abakaliki smallpox data.}\label{tab:tabEpi}
\end{table*}

\begin{table}[t]
\centering
\small
	\begin{tabular}{@{}l lll lll@{}}
         \toprule
Filter  & $N_{x}$ &  CPU & \multicolumn{4}{l}{Bias (RMSE)}  \\
\cmidrule(l){4-7}
	&         &            & $\widehat{E}(\log(c_1)|y_{1:T})$ & $\widehat{E}(\log(c_2)|y_{1:T})$ & $\widehat{SD}(\log(c_1)|y_{1:T})$ & $\widehat{SD}(\log(c_2)|y_{1:T})$\\   
\midrule
Bootstrap           & 404            &175 &0.068 (0.022) & \phantom{-}0.017 (0.023)  & -0.026 (0.012) & -0.011 (0.017) \\
Alive               & {\phantom{0}77}&256 &0.033 (0.026) & -0.006 (0.035) & -0.020 (0.014) & -0.008 (0.022) \\
Auxiliary (meth. 1) & \phantom{0}81  &\phantom{0}45  &0.041 (0.024) &-0.024 (0.028) & -0.024 (0.014) & -0.010 (0.016)\\
Auxiliary (meth. 2) & 174            &162 &0.037 (0.028) &\phantom{-}0.040 (0.034) & -0.022 (0.019) &-0.024 (0.020)  \\
\bottomrule
\end{tabular}
	\caption{SIR epidemic model (Abakaliki data). $N_{x}$ at time $T$, CPU time (in seconds), bias (and RMSE in parentheses) of estimators 
of the posterior expectations ${E}(\log(c_1)|y_{1:T})$, ${E}(\log(c_2)|y_{1:T})$ and standard deviations ${SD}(\log(c_1)|y_{1:T})$, ${SD}(\log(c_2)|y_{1:T})$. All results are obtained by averaging over 100 runs of each SMC$^2$ scheme.}\label{tab:tabEpi2}
\end{table}

\begin{figure}[t]
\centering
\psfragscanon
\psfrag{Time}[][][0.7][0]{Time}
\psfrag{ARate}[][][0.7][0]{Acc. rate}
\psfrag{ESS}[][][0.7][0]{ESS}
\psfrag{Nx}[][][0.7][0]{$N_x$}
\includegraphics[width=5.5cm,height=16cm,angle=-90]{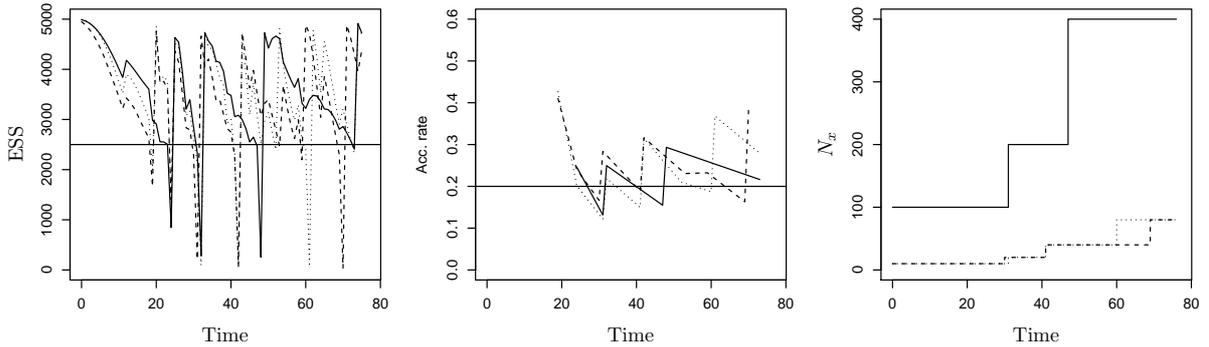}
\caption{SIR epidemic model (Abakaliki data). Left panel: Effective sample size (ESS) against time. Middle panel: 
Acceptance rate against time. Right panel: Number of state particles $N_{x}$ against time. Horizontal 
lines indicate the thresholds at which resampling and doubling of $N_x$ take place.  All results are based 
on a single typical run of an SMC$^2$ scheme using the bootstrap (solid line), alive (dashed line) and auxiliary method 1 (dotted line) particle filters. 
Auxiliary method 2 is omitted for ease of exposition.}
\label{fig:epi}
\end{figure}

\begin{figure}[t]
\centering
\psfragscanon
\psfrag{Time}[][][0.7][0]{Time}
\psfrag{log(c1)}[][][0.7][0]{$\log(c_1)$}
\psfrag{log(c2)}[][][0.7][0]{$\log(c_2)$}
\psfrag{log(c1c2)}[][][0.7][0]{$\log(c_1/c_2)$}
\includegraphics[width=5.5cm,height=16cm,angle=-90]{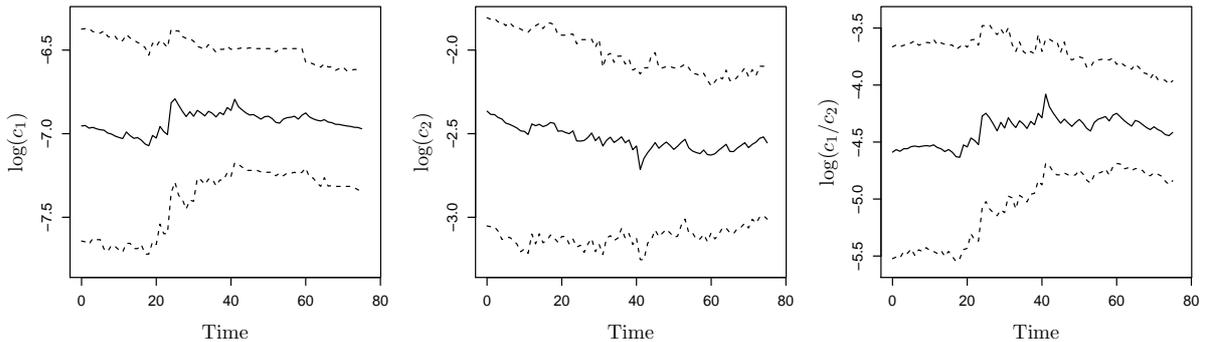}
\caption{SIR epidemic model (Abakaliki data). Marginal posterior mean (solid line) and $95\%$ credible interval (dashed lines) for 
$\log(c_1)$ (left), $\log(c_2)$ (middle) and $\log(c_1/c_2)$ (right) based on the output of 
the auxiliary SMC$^2$ scheme.}
\label{fig:epi1.5}
\end{figure}

\begin{figure}[t]
\centering
\psfragscanon
\psfrag{density}[][][0.7][0]{Density}
\psfrag{log(c1)}[][][0.7][0]{$\log(c_1)$}
\psfrag{log(c2)}[][][0.7][0]{$\log(c_2)$}
\includegraphics[width=5.5cm,height=16cm,angle=-90]{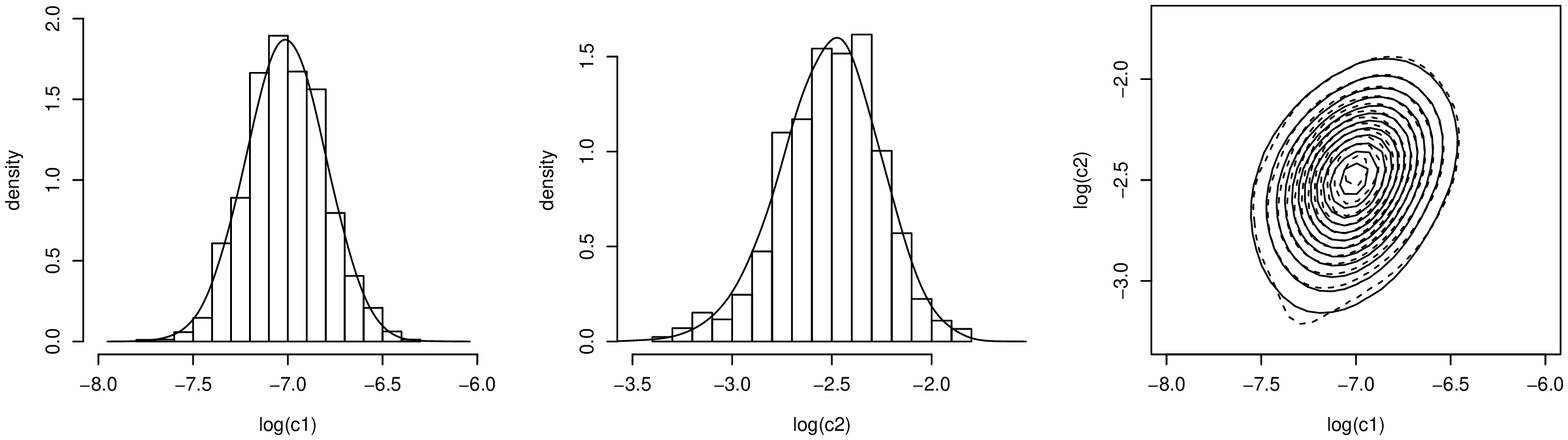}
\caption{SIR epidemic model (Abakaliki data). Left and middle panels: marginal posterior distributions based on the output of 
the auxiliary SMC$^2$ scheme (histograms) and pMCMC scheme (kernel density estimates). Right panel: Contour plot of the joint posterior from 
the output of the auxiliary SMC$^2$ scheme (dashed lines) and pMCMC (solid lines).}
\label{fig:epi2}
\end{figure}

We followed \cite{fearnhead2004} by taking independent Gamma priors 
so that $c_{1} \sim Ga(10,10^4)$ and $c_{2}\sim Ga(10,10^2)$ \emph{a priori}, where $Ga(a,b)$ denotes a Gamma distribution with shape $a$ and 
rate $b$. We applied three different SMC$^2$ schemes based on the bootstrap, alive and auxiliary 
(with propagation methods 1 and 2) particle filters. In each case we took $N_{c}=5000$, an ESS-threshold of $\gamma=50\%$ 
and an initial number of state particles of $N_{x}=10$, except when using the bootstrap filter which 
required $N_{x}=100$ initially, to give output comparable to the other methods, in terms of accuracy (-- see further discussion below). 
The value of $N_x$ was doubled if the acceptance rate calculated in the resample-move step fell below $20\%$. 

Table~\ref{tab:tabEpi2} and Figures~\ref{fig:epi}--\ref{fig:epi2} summarise the output 
of each SMC$^2$ scheme. We compare the accuracy of each scheme by reporting bias and root mean 
square error (RMSE) of the estimators of the marginal posterior means and standard deviations of 
$\log(c_1)$ and $\log(c_2)$. These quantities are reported in Table~\ref{tab:tabEpi2} and were obtained 
by performing 100 independent runs of each scheme and comparing the aforementioned posterior estimators 
to reference values, obtained from a long run ($3\times 10^6$ iterations) of particle MCMC (pMCMC). For the 
pMCMC run, we used the auxiliary particle filter 
driven scheme of \cite{GoliWilk15} which uses Algorithm~1 and propagation method~1 at each MCMC iteration 
to compute $\hat{p}(y_{1:T}|c^*)$ for a proposed value $c^*$. A comparison of SMC$^2$ and pMCMC is given 
in Section~\ref{MCMCcomp}.    

Inspection of Table~\ref{tab:tabEpi2} shows that all schemes give generally comparable output in terms 
of bias and RMSE, although we found that the bootstrap implementation was particularly 
sensitive to the initial choice of $N_x$, with relatively low values leading to noticeable biases in the 
marginal posterior mean estimators. Using 100 initial state particles seemed to alleviate this problem. We 
therefore use CPU cost as a proxy for overall efficiency. Interestingly, the alive SMC$^2$ scheme performs 
poorly in terms of CPU cost, despite requiring the smallest number of state particles. As can be seen 
from Figure~\ref{fig:epi} (left panel), alive SMC$^2$ maintains a high effective sample size (ESS), 
rarely falling below the threshold that would trigger the resample-move step. In spite of this desirable behaviour, 
the scheme requires repeatedly forward simulating the process at each time-point to obtain $N_x$ matches, 
resulting in a CPU cost that is almost 1.5 times larger than that obtained for the bootstrap driven scheme. 
Both auxiliary schemes outperform the bootstrap implementation, with method 1 by a factor of 3.9 
in terms of CPU cost. Finally, we note that the SMC$^2$ scheme allows for sequential learning of the 
rate constants as well as the basic reproduction number $R_0=c_1/c_2$ -- see Figure~\ref{fig:epi1.5} 
showing marginal posterior means and $95\%$ credible intervals against time. Figure~\ref{fig:epi2} 
compares the output of an SMC$^2$ scheme with the output of a long run of pMCMC and demonstrates that 
accurate fully Bayesian inferences about the parameters are possible, 
even when using relatively few parameter particles.

\subsubsection{Comparison with MCMC}\label{MCMCcomp}

Here, we assess the utility of the auxiliary particle filter (method 1) driven SMC$^2$ scheme 
as an offline inference scheme by comparing its performance to that of two competing MCMC schemes, 
namely the particle MCMC scheme used by \cite{GoliWilk15} and a data augmentation
scheme first introduced by \citet{oneill1999} and \citet{gibson1998}.

As discussed earlier, the likelihood of the observed data (i.e. removal times) is challenging to compute. The reason is that one has to integrate out all the possible configurations of infection times that are consistent with the data; in other words, those that do not result in the epidemic ceasing before the last removal time. One way to overcome this issue is to introduce the unobserved infection times as additional variables which will allow us to compute an augmented likelihood. Combining the augmented likelihood with prior distributions on the infection rate ($c_1$) and removal rate ($c_2$), we can then explore the joint posterior density of the infection times, $c_1$ and $c_2$ using a data-augmented Markov Chain Monte Carlo scheme (DA-MCMC). 

A vanilla DA-MCMC algorithm consists of updating $c_1$, $c_2$ and the infection times from their corresponding full conditional (posterior) densities. It turns out that the full conditional densities for $c_1$ and $c_2$ have standard forms and can be updated using a Gibbs step; in fact, both full conditional densities are Gamma densities. The infection times are less straightforward to deal with because the the full conditional distribution of each infection time is not of a standard form. However, they can be updated by using a Metropolis-Hastings step. This is done by proposing a new infection time and accepting that proposed infection times with some probability determined by the Metropolis-Hastings ratio. In particular, a new infection time for the $j$th individual, $i_j^*$, is proposed by drawing $X \sim \mbox{Exp}(c_2)$ and setting $i_j^* = r_j - X$ where $r_j$ denotes the corresponding removal time of individual $j$.

\begin{figure}[t]
\centering
\psfragscanon
\psfrag{density}[][][0.7][0]{Density}
\psfrag{log(c1)}[][][0.7][0]{$\log(c_1)$}
\psfrag{log(c2)}[][][0.7][0]{$\log(c_2)$}
\includegraphics[width=5.5cm,height=16cm,angle=-90]{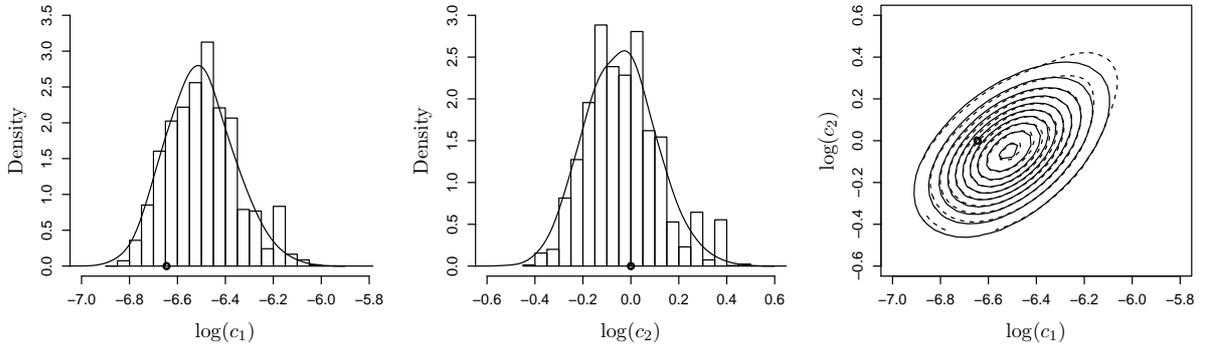}
\caption{SIR epidemic model (synthetic data). Left and middle panels: marginal posterior distributions based on the output of 
the auxiliary SMC$^2$ scheme (histograms) and pMCMC scheme (kernel density estimates). Right panel: Contour plot of the joint posterior from 
the output of the auxiliary SMC$^2$ scheme (dashed lines) and pMCMC (solid lines). The true values of $\log(c_1)$ and $\log(c_2)$ are indicated.}
\label{fig:epi4}
\end{figure}

To provide a challenging scenario, we assumed a fixed population size of $n=1000$, an infection rate of 
$c_1=0.0013$, a removal rate of $c_2=1$ and generated a synthetic data set consisting of 622 inter-removal times, equivalent to 622 measurements of $X_{1}+X_{2}$. For simplicity, we  assume that the initial condition $x_0=(n-1,1)'$ is known. 
We took vague Exponential $Exp(0.001)$ priors  for each rate constant and performed 50 runs of (auxiliary) SMC$^2$, pMCMC and MCMC-DA with the following settings.
\begin{enumerate}
\item \emph{SMC$^2$}. We took $N_{c}=5000$, an ESS-threshold of $\gamma=50\%$ 
and an initial number of state particles of $N_{x}=100$. The value of $N_x$ was doubled if the 
acceptance rate calculated in the resample-move step fell below $20\%$. Note that initialising 
with a sample from the vague prior would result in very few parameter particles consistent 
with the first observation. This problem can be alleviated by partitioning the interval $[0,1]$ 
into $m+1$ equally spaced intermediate time points and targeting the tempered posteriors 
$p(c)p(y_{1}|c)^{i/m}$, $i=0,1,\ldots,m$. We adopted an alternative solution and performed $10000$ pMCMC iterations using 
the first 10 observations (with $N_x=100$), thinned by a factor of 2 and then ran SMC$^2$ for 
the remaining 612 observations, having initialised with the pMCMC output. 
     
\item \emph{pMCMC}. Following the practical advice of \cite{sherlock2015}, the number of state particles was chosen 
so that the variance of the estimator of the log-posterior at the posterior median (obtained from a pilot run) was around 
2. This gave $N_x=1200$. A random walk proposal was used for the log-parameters with the variance of the Gaussian 
innovations taken to be $\widehat{Var}(\log(c)|y_{1:T})$ (estimated from a pilot run) and scaled to give an acceptance rate of 
around $10\%-15\%$. The same pilot run was used to obtain the estimate $\widehat{E}(\log(c)|y_{1:T})$ and the main monitoring runs 
were initialised using this value.   

\item \emph{MCMC-DA}.  It has been illustrated that in practice \citep{Kypraios07}, if the infection-time update step is repeated several times in each iteration of the MCMC algorithm then mixing can improve substantially. Denote the fraction of infection times to be update in each MCMC step by $\delta$. After running  a number of short pilot runs with $\delta\in\{0.1,0.2,0.3,0.4,\linebreak[1]0.5,0.6,0.7\}$, we found that $\delta=0.5$ was optimal in terms of minimising autocorrelation time (defined below). The main monitoring runs then used $\delta=0.5$ and were initialised with the same values used for the pMCMC runs.
\end{enumerate}
Note that the number of iterations of pMCMC and MCMC-DA performed for the 50 runs was determined by the CPU cost of each run of SMC$^2$. 
Consequently, all results are reported for the same computational budget. The results are summarised in Table~\ref{tab:tabEpi4} and 
Figure~\ref{fig:epi4}. From the latter, it is clear that the output of SMC$^2$ is comparable with that of pMCMC. The two competing 
MCMC schemes can be directly compared by computing autocorrelation time (ACT), sometimes referred to as inefficiency, and can be interpreted 
as the factor by which the number of iterations ($n_{\textrm{iters}}$) should be multiplied, to obtain the same level of precision as using 
$n_{\textrm{iters}}$ iid posterior draws. The ACT for a particular series of parameter values is given by
\[
1+2\sum_{k=1}^{\infty}\rho_k
\]   
where $\rho_k$ is the autcorrelation function for the series at lag $k$. The ACT can be computed using the \verb+R+ package CODA \citep{Plummer06}.

The MCMC-DA scheme is relatively cheap, with the CPU budget affording runs of around $10^6$ iterations on average. 
By comparison, the pMCMC scheme typically used around $2.5 \times 10^4$ iterations. However, the mixing of MCMC-DA 
is very poor, due to the dependence between the parameter values and the imputed infection times. For pMCMC, a joint update of the parameters and 
latent infection times is used (thereby side-stepping the issue of high correlation between the two) and mixing is much improved. 
Consequently, for MCMC-DA, the maximum (over each parameter series) ACT is around 8 times larger than that for pMCMC (after matching iteration numbers). 
Not surprisingly, estimators of the marginal posterior means and standard deviations for the log rate constants based on MCMC-DA exhibit biases and root mean 
square errors that are significantly larger than those obtained for pMCMC. Using SMC$^2$ gives output comparable to that of pMCMC, with 
all biases within an order of magnitude of those for pMCMC, and all RMSE values within a factor of 3. Moreover, it should be noted that 
we are comparing against a pMCMC scheme with (close to) optimal settings obtained from pilot runs. SMC$^2$ requires minimal tuning 
by comparison, yet appears to be an effective offline inference tool in this example.        

\begin{table}[t]
\centering
\small
	\begin{tabular}{@{}l lll lll@{}}
         \toprule
Method  & $N_{x}$ &  mACT & \multicolumn{4}{l}{Bias (RMSE)}  \\
\cmidrule(l){4-7}
        &         &          & $\widehat{E}(\log(c_1)|y_{1:T})$ & $\widehat{E}(\log(c_2)|y_{1:T})$ & $\widehat{SD}(\log(c_1)|y_{1:T})$ & $\widehat{SD}(\log(c_2)|y_{1:T})$\\   
\midrule
SMC$^2$ & 1120& --            &-0.004 (0.017)            & -0.009 (0.016)            & -0.009 (0.009)            & -0.008 (0.009) \\
pMCMC   & 1200&\phantom{0}29  &\phantom{-}0.001  (0.006) &\phantom{-}0.001   (0.006) & -0.003 (0.003)            & -0.001 (0.003)\\
MCMC-DA & --  &232            &-0.006 (0.017)            &-0.002 (0.017)             & \phantom{-}0.036 (0.014)  & \phantom{-}0.037 (0.015)  \\
\bottomrule
\end{tabular}
      \caption{SIR epidemic model (synthetic data). $N_{x}$ at time $T$, maximum autocorrelation time (mACT), bias (and RMSE in parentheses) of estimators 
of the posterior expectations ${E}(\log(c_1)|y_{1:T})$, ${E}(\log(c_2)|y_{1:T})$ and standard deviations ${SD}(\log(c_1)|y_{1:T})$, ${SD}(\log(c_2)|y_{1:T})$. All results are obtained by averaging over 50 runs of each scheme. Note that mACT for MCMC-DA has been scaled to correspond to the average number of iterations of pMCMC.}\label{tab:tabEpi4}	
\end{table}

\subsection{Prokaryotic autoregulation}\label{prok}

Using the model of prokaryotic autoregulation described in Section~\ref{ex2}, we 
simulated two synthetic data sets (denoted $\mathcal{D}_1$ and $\mathcal{D}_2$) 
consisting of 101 observations at integer times on \textsf{RNA} and total protein 
counts, $\textsf{P}+2\textsf{P}_2$, so that \textsf{DNA}, \textsf{P} and $\textsf{P}_2$ 
are not observed exactly. Moreover, we corrupt the 
observations by adding independent, zero-mean Gaussian innovations to each count. The 
components making up the observation in (\ref{obs_eq}) are 
\[
P'=\begin{pmatrix}
1 & 0 & 0 & 0\\
0 & 1 & 2 & 0
\end{pmatrix}, \qquad \Sigma=\begin{pmatrix}
\sigma_1^2 & 0\\
0 & \sigma_2^2
\end{pmatrix}.
\] 
To assess the effect of measurement error, we fix $\sigma_2=1$ and take 
$\sigma_{1}=1$ for data set $\mathcal{D}_1$ and $\sigma_1=0.1$ for 
$\mathcal{D}_2$. Following \cite{GoliWilk05}, the rate constants used to 
generate the data were
\[
c=(0.1,0.7,0.35,0.2,0.1,0.9,0.3,0.1).
\]
We assume that the initial condition $x_0=(8,8,8,5)'$, the measurement 
error variances and the rate constants of the reversible dimerisation 
reactions ($c_5$ and $c_6$) are known leaving 6 parameters as the object 
of inference. 

We took independent Gamma $Ga(1,0.5)$ priors for each rate constant and 
applied SMC$^2$ schemes based on the bootstrap and auxiliary (with propagation method 1) 
particle filters. In each case we took $N_{c}=5000$, an ESS-threshold of $\gamma=50\%$ 
and an initial number of state particles of $N_{x}=50$. The value of $N_x$ was doubled if the 
acceptance rate calculated in the resample-move step fell below $20\%$.

\begin{table}[t]
\centering
\small
	\begin{tabular}{@{}l lll lll@{}}
         \toprule
\quad Filter  & $N_{x}$ &  CPU & \multicolumn{4}{l}{Bias (RMSE)}  \\
\cmidrule(l){4-7}		
        &         &          & $\widehat{E}(\log(c_1)|y_{1:T})$ & $\widehat{E}(\log(c_2)|y_{1:T})$ & $\widehat{SD}(\log(c_1)|y_{1:T})$ & $\widehat{SD}(\log(c_2)|y_{1:T})$\\   
\midrule
\multicolumn{7}{@{}l}{$\mathcal{D}_{1}$ ($\sigma_{1}=1$, $\sigma_{2}=1$)}\\
\quad Bootstrap &2688           &495      &-0.281 (0.051) &-0.066 (0.053) &-0.094 (0.030) &-0.077 (0.036) \\
\quad Auxiliary &\phantom{0}564 &242      &-0.129 (0.082) &-0.027 (0.067) &-0.098 (0.064) &-0.084 (0.033) \\
\\
\multicolumn{7}{@{}l}{$\mathcal{D}_{2}$ ($\sigma_{1}=0.1$, $\sigma_{2}=1$)}\\  
\quad Bootstrap &8000 &1905           &-0.011 (0.082) &-0.088 (0.063) &-0.063 (0.056) &-0.063 (0.046)  \\
\quad Auxiliary &1120 &\phantom{0}474 &-0.047 (0.048) &-0.029 (0.047) &-0.079 (0.020) &-0.058 (0.024) \\
\bottomrule
\end{tabular}
      \caption{Prokaryotic autoregulation. $N_{x}$ at time $T$, CPU time (in minutes), bias (and RMSE in parentheses) of estimators 
of the posterior expectations ${E}(\log(c_1)|y_{1:T})$, ${E}(\log(c_2)|y_{1:T})$ and standard deviations ${SD}(\log(c_1)|y_{1:T})$, ${SD}(\log(c_2)|y_{1:T})$. 
All results are obtained by averaging over 50 runs of each scheme.}\label{tab:tabAR}	
\end{table}

\begin{figure}[t]
\centering
\psfragscanon
\psfrag{density}[][][0.7][0]{Density}
\psfrag{log(c1)}[][][0.7][0]{$\log(c_1)$}
\psfrag{log(c2)}[][][0.7][0]{$\log(c_2)$}
\psfrag{log(c3)}[][][0.7][0]{$\log(c_3)$}
\psfrag{log(c4)}[][][0.7][0]{$\log(c_4)$}
\psfrag{log(c7)}[][][0.7][0]{$\log(c_7)$}
\psfrag{log(c8)}[][][0.7][0]{$\log(c_8)$}
\includegraphics[width=5.5cm,height=16cm,angle=-90]{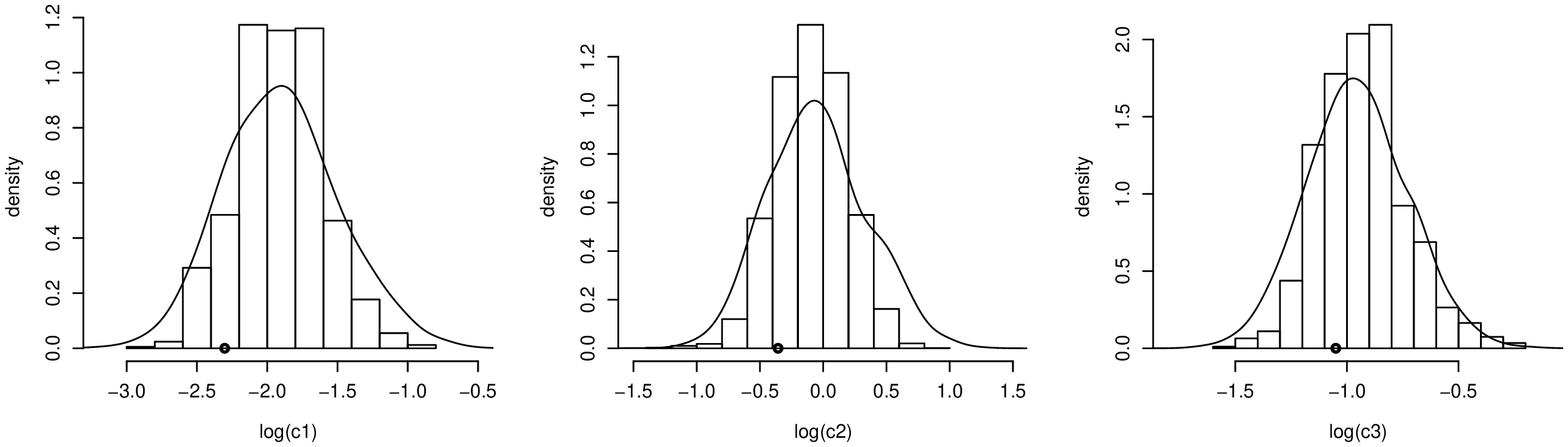}
\includegraphics[width=5.5cm,height=16cm,angle=-90]{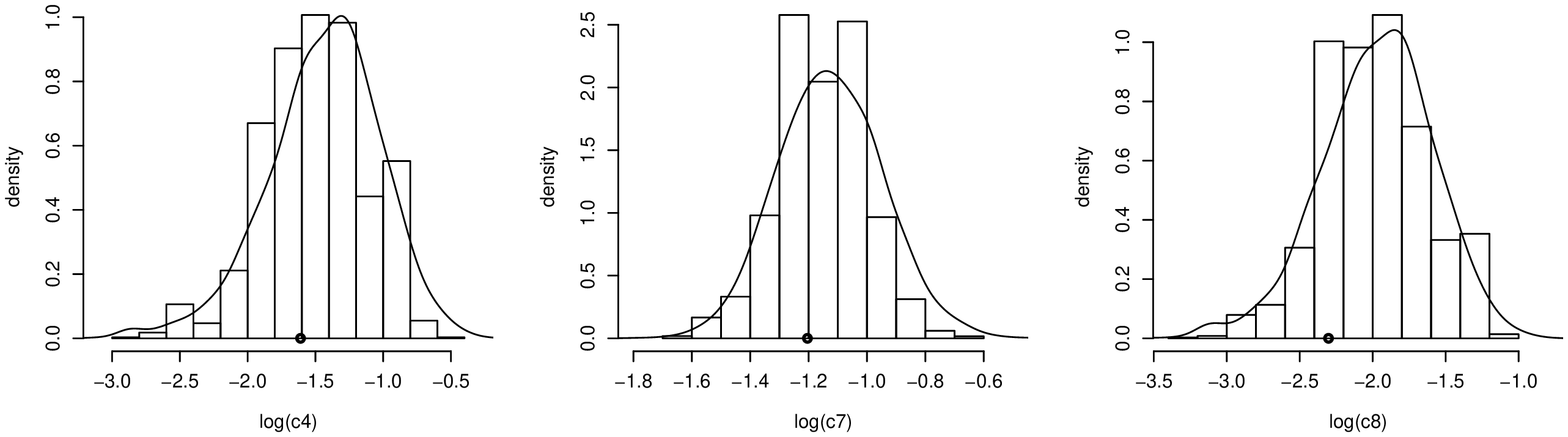}
\caption{Prokaryotic autoregulation (synthetic data set $\mathcal{D}_{2}$). Marginal posterior distributions based on the output of 
the auxiliary SMC$^2$ scheme (histograms) and pMCMC scheme (kernel density estimates). The true values of the (log) rate constants are indicated.}
\label{fig:AR0}
\end{figure}

\begin{figure}[t]
\centering
\psfragscanon
\psfrag{Time}[][][0.7][0]{Time}
\psfrag{ARate}[][][0.7][0]{Acc. rate}
\psfrag{ESS}[][][0.7][0]{ESS}
\psfrag{Nx}[][][0.7][0]{$N_x$}
\includegraphics[width=5.5cm,height=16cm,angle=-90]{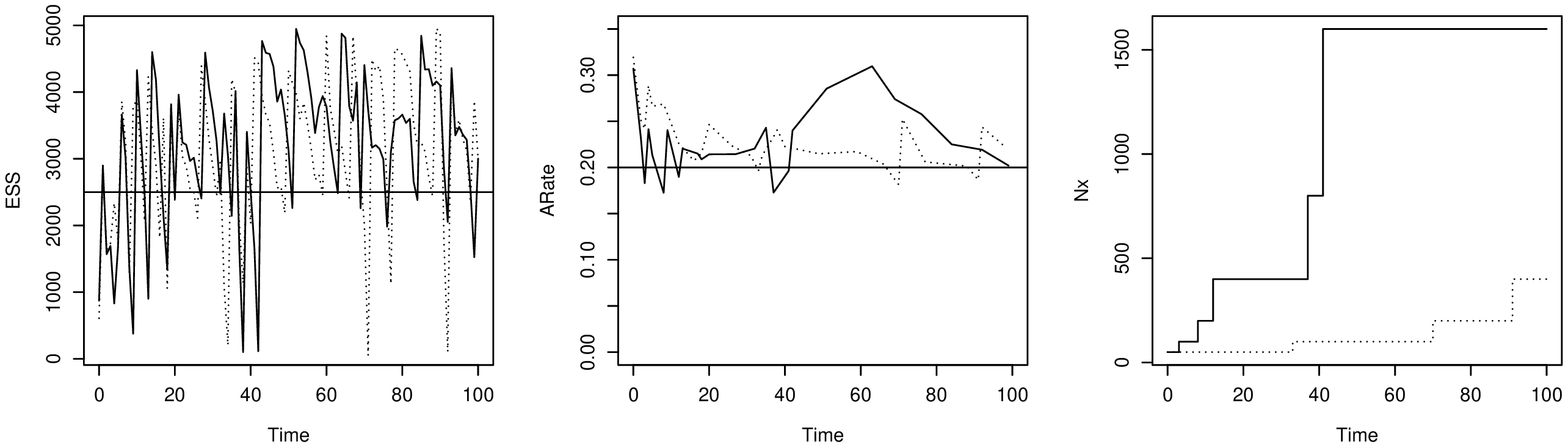}
\includegraphics[width=5.5cm,height=16cm,angle=-90]{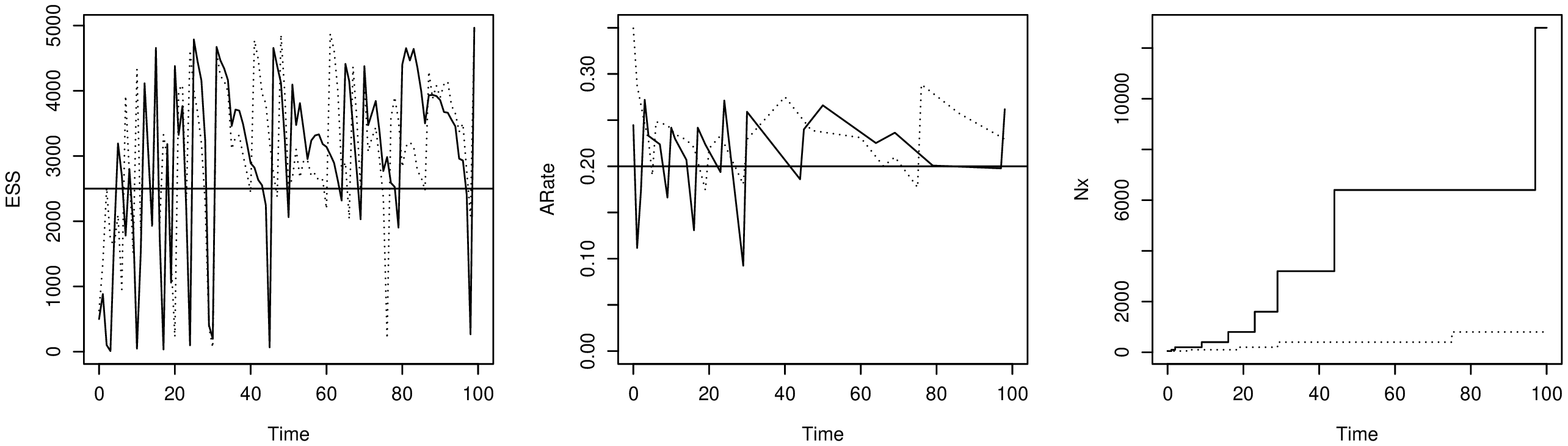}
\caption{Prokaryotic autoregulation using synthetic data sets $\mathcal{D}_{1}$ (top panel) and 
$\mathcal{D}_{2}$ (bottom panel). Left panel: Effective sample size (ESS) against time. Middle panel: 
Acceptance rate against time. Right panel: Number of state particles $N_{x}$ against time. Horizontal 
lines indicate the thresholds at which resampling and doubling of $N_x$ take place.  All results are based 
on a single typical run of an SMC$^2$ scheme using the bootstrap (solid line) and auxiliary (dotted line) particle filters.}
\label{fig:AR}
\end{figure}

Figure~\ref{fig:AR0} shows marginal posteriors based on the output of auxiliary SMC$^2$ and a long run of 
pMCMC. We note that even with 6 unknown parameters, the SMC$^2$ scheme gives accurate inferences despite using relatively 
few parameter particles. Table~\ref{tab:tabAR} and Figure~\ref{fig:AR} summarise the output 
of each SMC$^2$ scheme. We again compare the accuracy of each scheme via bias and RMSE 
of the estimators of the marginal posterior means and standard deviations of the (log) rate 
constants. Bias and RMSE were computed by comparing estimators based on 50 runs of each SMC$^2$ scheme with reference 
values obtained from a long run of pMCMC (with $5\times 10^5$ iterations). Table~\ref{tab:tabAR} displays these 
quantities for $\log(c_1)$ and $\log(c_2)$ corresponding to the reversible dimer 
binding and unbinding reactions. Similar results (not shown) are obtained for the remaining unknown rate constants. 
Both the bootstrap and auxiliary particle filter driven schemes give comparable bias and RMSE values 
and we therefore compare their overall performance using CPU cost. Not surprisingly, as the measurement error is reduced, both schemes require 
increased numbers of state particles, $N_x$, although the relative increase is much smaller when using auxiliary SMC$^2$. 
Consequently, for data set $\mathcal{D}_{1}$ ($\sigma_1=1$), auxiliary SMC$^2$ outperforms bootstrap SMC$^2$ in terms 
of CPU time by around a factor of 2. This increases to a factor of around 4 for data set 
$\mathcal{D}_{2}$ ($\sigma_1=0.1$).

\section{Discussion}

Performing fully Bayesian inference for the rate constants governing complex 
stochastic kinetic models necessitates the use of computationally intensive 
Markov chain Monte Carlo (MCMC) methods. The intractability of the observed 
data likelihood further complicates matters, and is usually dealt with 
through the use of data augmentation or by replacing the intractable 
likelihood by an unbiased estimate. Careful implementation of the latter 
results in a pseudo-marginal Metropolis-Hastings scheme, and, when using 
a particle filter to obtain likelihood estimates, the algorithm may be 
referred to as particle MCMC (pMCMC). However, such methods often 
require careful tuning and initialisation and do not allow for efficient 
sequential learning of the parameters (and latent states). 

We have therefore focused on a recently proposed SMC$^2$ scheme, which can 
be seen as the pseudo-marginal analogue of the iterated batch importance 
sampling (IBIS) scheme \citep{chopin2002}, and allows sequential 
learning of the parameters of interest. The simplest implementation 
uses a bootstrap particle filter both to compute observed-data 
likelihood increments and drive a rejuvenation step (so called resample move) 
where all parameter particles are mutated through a pMCMC kernel. This 
simple implementation is appealing -- for example, only the ability to 
evaluate the density associated with the observation equation, and 
generate forward realisations from the Markov jump process is required. 
However, this `likelihood-free' implementation is likely to be extremely 
inefficient when observations are informative, e.g. when there is relatively 
little measurement error compared to intrinsic stochasticity. We eschew 
the simplest implementation in favour of an SMC$^2$ scheme that is driven 
by an auxiliary particle filter (APF). That is, the APF is used both to 
estimate the observed-data likelihood contributions and drive the resample-move 
step. We compared this approach using two applications: an SIR epidemic 
model fitted to real data and a simple model of prokaryotic autoregulation 
fitted to synthetic data. 

We find that the proposed approach offers significant gains in computational efficiency 
relative to the bootstrap filter driven implementation, whilst still maintaining 
an accurate particle representation of the full posterior. The computational gains 
are amplified when intrinsic stochasticity dominates external noise (e.g. measurement error). Use of an 
appropriate propagation mechanism is crucial in this case, since the probability 
of generating an (unconditioned) realisation of the latent jump process that is 
consistent with the next observation, diminishes as either the observation variance 
decreases or the number of observed components increases.

Using synthetic data and the SIR epidemic model, we also compared the 
efficiency of SMC$^2$ with two competing MCMC schemes, namely the APF driven 
particle MCMC scheme of \cite{GoliWilk15} and a ubiquitously applied data 
augmentation (DA) scheme \citep{oneill1999,gibson1998}. We find that the DA scheme suffers intolerably 
poor mixing due to dependence between the latent infection times and the static 
parameters (see also \cite{mckinley2014}). The pMCMC scheme, which can be seen as the pseudo-marginal analogue of 
an idealised marginal scheme, offers over an order of magnitude increase in terms 
of overall efficiency (as measured by autocorrelation time for a fixed computational budget) 
over DA. The APF driven SMC$^2$ scheme gives comparable output to that of pMCMC 
in terms of accuracy (as measured by bias and root mean squared error of key 
posterior summaries). However, we stress again that unlike pMCMC, SMC$^2$ is 
simple to initialise, avoids the need for tedious pilot runs, performs sequential 
learning of the parameters of interest and allows for a computationally efficient 
estimator of the model evidence. Although not persued here, model selection is an 
important problem within the stochastic kinetic framework (see e.g. \cite{drovandi2016} 
and the references therein for recent discussions).

\subsection{Use of other particle filters}

The development of an auxiliary particle filter driven SMC$^2$ scheme 
as considered in this paper is possible due to the tractability of the 
complete data likelihood $p(x_{(t-1,t]}|x_{t-1},c)$ for each observation 
time $t$. This tractability may permit the use of other particle filtering 
strategies. For example, particle Gibbs with ancestor sampling \citep{lindsten14} 
allows for efficient sampling of state trajectories, and could be used in the 
rejuvenation step in SMC$^2$.  Recent work by \cite{guarniero16} combines 
ideas underpinning the twisted particle filter of \cite{whitely14} and 
the APF to give the iterated APF (iAPF). The algorithm approximates an idealized 
particle filter where observed data likelihood estimates have zero variance. 
Consequently, use of this approach in an SMC$^2$ requires further attention, although 
it would appear the iAPF algorithm is at present limited to a class of state space models 
with conjugate latent processes. Its utility within the SKM framework is therefore 
less clear.

\subsection{Further considerations}

This work can be directly extended in a number of ways. In our application of the APF, 
we assumed a constant preweight for each parameter particle. Devising a preweight that 
is both computationally cheap and accurate remains of interest. In addition, 
the best performing propagation method is derived using a linear Gaussian 
approximation to the number of reaction events in an interval of interest, 
conditional on the next observation. Improvements to this construct that 
allow for a more accurate approximation of the intractable conditioned process 
are the subject of ongoing work. Although not considered here, the 
SMC$^2$ scheme appears to be particularly amenable to parallelization over parameter 
particles, since observed data likelihood estimates can be computed separately for 
each parameter value. The use of parallel resampling algorithms 
\citep{murray16} also merits further attention, to allow full use of modern computational architectures. 
Finally, we note that the resample-move step 
may benefit from recent work on correlated pseudo-marginal schemes \citep{dahlin2015,
deligiannidis2016}.

\

\noindent\textbf{Acknowledgements} The authors would like to thank the
associate editor and anonymous referee for their suggestions for
improving this paper.

\bibliographystyle{apalike}
\bibliography{bridgebib}   

\appendix

\section{Appendix}


\subsection{Alive SMC$^2$}\label{alive}

Consider the case that $\Sigma=0$ so that (a subset of) the components of $X_t$ are observed 
without error. Running a bootstrap particle filter in this scenario is likely to be problematic, 
since only trajectories which match the observation at each time $t$ will be assigned a non-zero weight. 
To circumvent this problem, \cite{drovandi2016} use the alive particle filter of \cite{delmoral2015} 
inside the SMC$^2$ scheme. Essentially, at time $t$, $N_x$ particles from time $t-1$ are resampled and 
propagated forward (using Gillespie's direct method) until $N_{x}+1$ matches are obtained 
(where a match has $x_t=y_t$). This approach can be repeated for each time point and 
an unbiased estimator of the observed-data likelihood $p(y_{1:t}|c)$ can then be obtained \citep{delmoral2015}.

The alive particle filter is described in Algorithm~\ref{alivePF}. Note that we have assumed for 
simplicity that $x_1$ is known, although a more general scenario with uncertain $x_1$ is easily accommodated by 
augmenting $c$ to include the unobserved components of $x_1$. The alive SMC$^2$ algorithm is obtained by 
running the alive particle filter in steps 2(a) and 3(b) of Algorithm~\ref{smc2alg}. 

\begin{algorithm}[t]
\caption{Alive particle filter}\label{alivePF}
\begin{enumerate}
\item Initialisation ($t=1$). Set $\hat{p}(y_{1}|c)=1$ and $x_{1}^{i}=y_{1}$, $i=1,\ldots,N_x$. For $t=2,3,\ldots ,T$:
\item For $i=1,2,\ldots$ until $i=n_{t}$ is reached such that $\sum_{j=1}^{i}I(x_{t}^{i},y_{t})=N_{x}+1$:
\begin{itemize}
\item[(a)] Sample $a_{t-1}^{i}$ uniformly from $\{1,\ldots,N_x\}$.
\item[(b)] Sample $x_{(t-1,t]}^{i}\sim p\big(\cdot|x_{t-1}^{a_{t-1}^{i}},c\big)$ using Gillespie's direct method.
\item[(c)] Calculate
\[
I(x_{t}^{i},y_{t})=\left\{\begin{array}{ll}
1, & x_{t}^{i}=y_{t}\\
0, & \textrm{otherwise}
\end{array}\right.
\]
\end{itemize}
\item Compute the current estimate of observed-data likelihood $\hat{p}(y_{1:t}|c)=\hat{p}(y_{1:t-1}|c)\hat{p}(y_{t}|y_{1:t-1},c)$ where
\[
\hat{p}(y_{t}|y_{1:t-1},c)=\frac{N_x}{n_t -1} .
\]
\end{enumerate}
\end{algorithm}

\end{document}